\documentclass{article}
\usepackage{natbib}

\title{PVLAS Experiment: Some Astrophysical Consequences}
\author{Gnedin Yu. N.\thanks{E-mail: gnedin@gao.spb.ru},
Piotrovich M. Yu., Natsvlishvili T. M.\\ Central Astronomical
Observatory at Pulkovo,\\ Saint-Petersburg, Russia.}

\begin{document}

\maketitle

\begin{abstract}
The birefringent effects of photon-pseudoscalar boson (Goldstone)
particle mixing in intergalactic magnetic field are calculated for
cosmological objects. We use the recent results of PVLAS
collaboration that reported recently the observation of a rotation
of the polarization plane of light propagating through a
transverse static magnetic field. Such results which were
interpretated as arising due to conversion of photon into
pseudoscalar with coupling strength $g_{a\gamma}\sim 4\times
10^{-6} GeV^{-1}$ allows us to estimate the intergalactic magnetic
field magnitude as $\sim 10^{-16} G$ based on Hatsemekers et al.
observations of extreme-scale alignments of quasar polarization
vectors. We have analyzed some additional results of astronomical
observations that could be explained by axion interpretation of
the PVLAS data: a sharp steepening of the QSO continuum short ward
of $\simeq 1100$\AA , observed circular polarization of AGNs and
QSOs, discrepancy between observed intrinsic polarization of stars
in the Local Bubble and stellar spectral classification. The
observed polarization of stars in the Local Bubble can't be
explained by interstellar origin.

\end{abstract}

\section{Introduction}

Recently \citet{b1} have reported in the PVLAS experiment a
rotation of polarization of light in vacuum in the presence of a
transverse magnetic field. The PVLAS experiment \citep{b2}
operates an ellipsometer acting with a superconducting dipole
magnet that can measure ellipticity and the polarization plane
rotation induced by the magnetic field onto linearly polarized
laser light. The sensitivity of the instrument is about $10^{-7}
rad Hz^{-1/2}$.

The experimental result of \citet{b1} proves the effect of light
polarization rotation in vacuum in the transverse magnetic field.
The averaged measured rotation is $(3.9\pm 0.5)\times 10^{-12}
rad/pass$, at 5T magnetic field strength with 44000 passes through
1m long magnet. The probable interpretation that was presented by
\citet{b1} is the magnetic conversion process of photons into
light, neutral, pseudoscalar particles (see, for example, the
reviews \citet{b3,b4,b5,b6}). The birefringent effects in vacuum
in the presence of a magnetic field may be produced by mixing
between the Nambu-Goldstone bosons (NGB) and photons. There exist
various ideas in favor of the existence of light and ultra-light
pseudoscalar particles beyond the Standard Model. These particles
are remained undetected due to their weak coupling to ordinary
matter including the electromagnetic field. The NGB and axion were
postulated to explain why the strong interaction conserves CP
(combined particle-antiparticle exchange and space conversion) in
spite of the fact that the weak interactions violate this
symmetry. Axion is the very popular candidate into the dark matter
(DM) particles. The typical values of standard axion mass lie in
the range $m_a \approx 10^{-5}\div 1eV$.

Cosmic NGBs and axions are produced in various Big Bang scenarios
(see reviews \citet{b7,b8,b9,b10}). Typically these are inflation
and cosmic strings. In cosmic strings scenario with the high $T\gg
T_{QSD}$ temperatures massless bosons are produced. For low $T\ll
T_{QSD}$ temperatures NGBs acquire a mass. In inflation models the
nonmassless particles are presumably produced with a mass scale
lying in $\mu eV$ range. Nevertheless, the chaotic inflation
models with high values of the Hubble parameter during inflation
$H_1\sim 10^{13}\div 10^{14} GeV$ provide the very low mass range
$m_a\leq 10^{-16} eV$, that is practically massless bosons.

NGBs and axions can be detected through their coupling to photons.
Nonmassless axion can decay into two photons. This $a\rightarrow
2\gamma$ coupling arises due to two different decay mechanisms:
through axion-pion mixing and via the electromagnetic (EM) anomaly
of PQ symmetry. But there is more effective process of coupling
between NGBs (axions) and photons in a magnetic field (Primakoff
process). This process can lead to photon production with energy
compared to the total boson energy. There exists certainly the
inverse process of transformation (conversion) of a photon into
NGB or axion in a magnetic field. These both processes are
effectively used for the search of the relic and stellar origin
axions \citep{b4,b6,b7,b8,b9,b10}.

The interpolations of pseudoscalars with Standard Model particles
are model dependent. The most strongest constraints come from
pseudoscalar coupling to photon: $g_{a\gamma}$. The corresponding
term of Lagrangian is
\begin{equation}
L_{int} = \frac{1}{4} g_{a\gamma} A F_{\mu\nu} \tilde{F}^{\mu\nu}
= g_{a\gamma} A \vec{E} \vec{B} \label{eq1}
\end{equation}
where $A$ is the pseudoscalar field.

The PVLAS experiment derives the mean value of coupling constant
as $g_{a\gamma}\approx 4\times 10^{-6} GeV^{-1}$ \citep{b1,b2}.

We consider the astrophysical consequences of the PVLAS
experiment, namely: (a) alignment of quasar polarization vectors
discovered recently by \citet{b11,b12}; (b) probable circular
polarization from distant ($z\sim 1$) quasars; (c) a sharp
steepening of the QSO continuum shortward of $\lambda\simeq
1100$\AA; (d) the origin of linear polarization of stars in the
Local Bubble.

\section{Photon-Goldstone Boson Mixing and Birefringent Effects in a Magnetic Field}

The probability of the magnetic conversion of photons into lowmass
bosons were calculated by \citet{b3} (they considered also the
conversion process of axions) and \citet{b13} (see also
\citet{b5}):
\begin{equation}
P(\gamma_{||}\longleftrightarrow a)=\frac{1}{1 +
x^2}\sin^2(\frac{1}{2}B_{\bot}g_{a\gamma}L\sqrt{1 + x^2})
\label{eq2}
\end{equation}
where
\begin{equation}
x = \bar{\varepsilon}\omega / B_{\bot}g_{a\gamma},\,\,
\bar{\varepsilon} = (\varepsilon - 1) / 2
\label{eq3}
\end{equation}
$\omega$ is the radiation frequency, $\varepsilon$ is the
dielectric function of a medium that the light is propagating
through, $B_{\bot}$ is the magnetic field component perpendicular
to the photon direction, $g_{a\gamma}$ is the coupling constant
between photons and pseudoscalars.

There are three main features of this probability function.

(a) Probability function has the oscillatory character.

(b) The sinus phase depends on the product of the magnetic
strength $B$, the size of the region $L$ where the magnetic field
is approximately homogeneous and the coupling constant of
pseudoscalar boson with photon $g_{a\gamma}$.

(c) The conversion process is very sensitive to the polarization
state of the photon since only a single polarization state with
the electric vector oscillating into the plane of the directions
of the magnetic field and the photon propagation is subject to the
conversion.

The conversion probability depends strongly on the dielectric
function of a medium. The most expression of this function was
given by \citet{b3} and \citet{b5}. Assuming that the medium is a
plasma with an electron density $N_e$ and a neutral gas (for
example, hydrogen) with density $N_H$, we find
\begin{equation}
1 - \varepsilon = \frac{\omega_p^2}{\omega^2} - 4\pi\beta N_H -
\frac{28\alpha^2}{45 m_e^4}B^2 - \frac{m_a^2}{\omega^2}
\label{eq4}
\end{equation}

The first term in (4) describes the contribution of the plasma
polarizability; the second one describes the contribution of a
neutral gas; the third one describes the contribution of, so
called, polarizability of vacuum in a magnetic field and is
important only for a plasma in the vicinity of neutron stars and
magnetic white dwarfs (see \citet{b14}). The last term in (4) is
the contribution of vacuum boson field. In a case of pure NGBs
$m_a = 0$.

In Eqs.(2)-(4) we use the Lorentz-Heaviside system of units with
$\hbar = c = 1$ and a fine structure constant $\alpha = e^2 / 4\pi
= 1 / 137$. In this system of units, the unit for the magnetic
field $B$ is $(eV)^2$, and the unit of length is $(eV)^{-1}$. Then
one gauss corresponds to $6.9\times 10^{-2} (eV)^2$ and 1cm
corresponds to $5\times 10^4 (eV)^{-1}$.

The various terms in (4) have opposite signs, so there is the
possibility that they would cancel out completely, and the Eq.(4)
and, therefore, $x$ would be equal to zero. This case corresponds
to the resonance in the probability function (2).

\citet{b15} has considered the case of resonance magnetic
conversion of photons into massless bosons from QSOs in the
regions of Damped $Ly_{\alpha}$ systems with a noticeable Faraday
rotation measure discovered by \citet{b16}. He interpreted with
this process the appearance of striking feature in polarized light
of some QSOs (see recent review by \citet{b17}). His case was a
compensation of first two terms in (4).

\citet{b3} and also \citet{b5} have discussed another possibility
of the resonance magnetic conversion process with compensation of
the first and last terms of Eq.(4). Naturally, such kind process
can take place only for a case of conversion into nonmassless
bosons (axions).

It appears the photon-boson mixing in a magnetic field provides
birefringent effects that are a rotation of polarization plane and
production of small amount of circular polarization.

Let remind that a plasma in a magnetic field possesses two
important magneto-optical effects: dichroism, i.e. the dependence
of extinction of light on its polarization state, and
birefringency, i.e. the difference of the refraction indexes or
phase velocities of polarized electromagnetic waves. In a strong
magnetic field of neutron stars and white dwarfs even the
electron-positron vacuum behaves as an anisotropic medium with the
birefringent properties \citep{b14}.

Photon-Goldstone boson (axion) mixing also yields birefringent
effects in a magnetic field because the change of parallel
polarization mode is produced via the conversion process of
photons into bosons (see Eq.(2)). Therefore the plane of
polarization will be rotated (oscillated) and ellipticity will be
acquired by a linear polarized beam propagated across magnetic
field lines in vacuum. Both effects were calculated by \citet{b18}
and by \citet{b3}.

In a result one can present the following expressions for the
rotation angle $\theta$ and circular polarization Stokes parameter
$V$:
\begin{equation}
\tan \theta (L) = \frac{1/2}{(1 +
x^2)}\sin^2(\frac{1}{2}B_{\bot}g_{a\gamma} L \sqrt{1 + x^2})
\label{eq5}
\end{equation}
\begin{equation}
P_C (L) = \frac{B_{\bot}}{2}g_{a\gamma}L\sqrt{1 + x^2}\left[1 -
\frac{\sin(\frac{B_{\bot}}{2}g_{a\gamma} L \sqrt{1 +
x^2})}{(\frac{B_{\bot}}{2}g_{a\gamma} L \sqrt{1 + x^2})}\right]
\label{eq6}
\end{equation}

In our case of a vacuum the weak mixing is most important so that
is a case $L\ll L_{Osc} = (B g_{a\gamma}\sqrt{1 + x^2})^{-1}$.

\begin{equation}
\theta (L)\approx \frac{1}{8} g_{a\gamma}^2 B_{\bot}^2 L^2
\label{eq7}
\end{equation}
\begin{equation}
P_C (L)\approx \frac{(B_{\bot} m_a)^2}{48\omega} g_{a\gamma}^2 L^3
\label{eq8}
\end{equation}

Eqs.(5)-(8) mean that ellipticity is acquired only for nonmassless
axions. Also there is a most important fact that the rotation
angle does not depend on photon frequency and boson mass.

\section{Cosmological Alignments of Quasar Optical Polarization Vectors}

Recently \citet{b11} and \citet{b12} have considered a sample of
170 optically polarized quasars with accurate linear polarization
measurements and discovered that quasar polarization vectors are
not randomly oriented over the sky as naturally expected. It
appeared that in some regions of three dimensional Universe (i.e.
in regions delimited in right ascension, declination and redshift)
the quasar polarization position angles are concentrated around
preferential directions, suggesting the existence of very
large-scale coherent orientation or alignment of quasar
polarization vectors. The existence of coherent orientations of
quasar polarization vectors have been later on confirmed in series
of works by \citet{b23} and \citet{b24} with use of a sample of
213 quasars. The final sample that was used for analysis includes
355 objects (see \citet{b12}). \citet{b12} used this sample of
quasars with significant optical polarization and using
complementary statistical methods they confirmed that quasar
polarization vectors are not randomly oriented over the sky with a
probability often in excess of 99.9\%. The polarization vectors
seem coherently oriented or aligned over huge ($\sim 1 Gpc$)
regions of the sky located in both low ($z\sim 0.5$) and high
($z\sim 1$) redshifts and looked characterized by different
preferred directions of the quasar polarization.

The linear dichroism of aligned interstellar dust grains in our
Galaxy produces linear polarization along the line of sight. This
polarization contaminates to some extent the quasar measured data
and may change their position angles. \citet{b25} have shown that
interstellar polarization has a little effect on the polarization
angle distribution of significantly polarized ($p_l\geq 0.6\%$)
quasars.

The interpretation of such large-scale alignment is difficult
within the commonly accepted cosmological models. Ongoing
theoretical works develop the idea that one might detect a
specific property of dark matter or dark energy. Preliminary
possible interpretations of the alignment effect have been
discussed by \citet{b11}, \citet{b23} and more recently by
\citet{b26,b24} and \citet{b12}. Since the alignments occur on
extremely large scales one must seek for global mechanisms acting
at cosmological scales.

From this point of view photon-pseudoscalar (ultra light axion or
axion-like pseudoscalar particle) mixing with a magnetic field
seems as a quite promising interpretation (see
\citet{b27,b5,b6,b28}) especially because many of the observed
properties of the alignment effect were qualitatively predicted.

Recently the exciting event have been taken place in the
photon-pseudoscalar mixing science. PVLAS Collaboration \citep{b2}
reported the experimental observation of a laser light
polarization rotation in vacuum in the presence of a transverse
magnetic field. They claimed that the average measured rotation is
$(3.9\pm 0.5)\times 10^{-12}\, rad/pass$, at $5T$ with 44000
passes through a 1m long magnet. Using Eq.(8) one can estimate the
value of the constant coupling of photon-axion mixing
$g_{a\gamma}$. We estimate its value as
\begin{equation}
g_{a\gamma}\approx 3.8\times 10^{-6} (GeV)^{-1}
\label{eq9}
\end{equation}

This value seems extremely higher than the corresponding value for
Peccei-Quin axion.

First all we may estimate the strength of the intergalactic (IG)
magnetic field using this value of the coupling constant
$g_{a\gamma}$, the characteristic size of alignment region $L\sim
1Gpc$ and the rotation rate of position angle $\Delta\theta\sim
30^0 (Gpc)^{-1}$ \citep{b12}. The Eq.(7) gives the following value
of the IG magnetic field:
\begin{equation}
B\approx 10^{-16} G
\label{eq10}
\end{equation}

This magnitude appears significantly less than the magnitude
predicted by some current works ($\leq 10^{-9}$ see, for instance,
\citet{b29,b30,b31,b32,b33,b34,b35,b36,b37}).

Nevertheless \citet{b38} presented a new model for generating
magnetic fields of cosmological interest. They have shown that the
photoionization process by photons from the first luminous sources
provides the magnetic field amplitudes as high as $2\times
10^{-16} G$. \citet{b42} discussed generation of magnetic field
from cosmological perturbations. They computed numerically the
magnitudes of the various contributions in the generation process
(three component plasma (electron, proton and photon) evolution,
the collision term between electrons and photons) and showed that
the amplitude of the produced magnetic field could be about $\sim
10^{-19} G$ at 10kpc co-moving scale at present.

\citet{b51} examined the generation of seed magnetic fields due to
the growth of cosmological perturbations. In the radiation era,
different rates of scattering from photons induce local
differences in the ion and electron density and velocity fields.
The currents due to the relative method of these fluids generate
magnetic fields on all cosmological scales. They estimated the
peak of a magnitude of these fields of $\sim 10^{-30} G$ at the
epoch of recombination. The major source of amplification of an
initial seed field comes from dynamo effect. A main problem is
connected with many mechanisms that produce quite weak seed fields
at times insufficiently early for dynamo amplifications. As an
example one should mention the Biermann mechanism that can produce
seed fields of order $\sim 10^{-19} G$, but only at redshift of
$z\sim 20$. It is remarkable that this magnitude is to be close to
our estimation (10) of intergalactic magnetic field strength.

\citet{b43} have discussed a new mechanism of generation of
intergalactic large-scale fields in colliding protogalactic clouds
and emerging protostellar clouds. Their mechanism is due to a
"shear-current" effect ("vorticity-current" effect) caused by the
large-scale shear motions of colliding clouds. Self-consistent
plasma-neutral gas simulations by \citet{b44} have shown that seed
magnetic field strengths $\leq 10^{-14} G$ arise in
self-gravitating protogalactic clouds of spatial scales of 100pc
during $7\times 10^6$ years.

Thus, we can conclude that the intergalactic magnetic field
magnitude is quite probably to be $\sim 10^{-16} G$, i.e. it is
keeping the value of early cosmological origin. Below we analyse
some additional observational consequences of positive PVLAS
experiment for astronomy.

\section{Cosmological Rotation of Polarization Plane in Radio Waves}

\citet{b19} have claimed that they found a systematic rotation of
the plane of polarization of electromagnetic radiation propagating
over cosmological distances. They have examined experimental data
on polarized radiation emitted by distant radio galaxies. After
extracting intergalactic Faraday rotation effect, the residual
rotation was found to follow a dipole rule. This residual rotation
appeared linear in the distance $L$ to the galaxy source and not
depending on a frequency of radiation. They claimed that this
effect could not be explained by uncertainties in subtracting
Faraday rotation. They determined a birefringence scale of order
$10^{25}h_{100}m$. Their claimed effect appears large, requiring
the plane of polarization from high redshift objects to be rotated
by as much as 3.0 rad - an extremely detectable signature.

However, \citet{b20} and \citet{b21} reexamined the same data and
argued that there is no statistically significant signal present.
\citet{b20} reported new optical data taken with the Keck
Telescope, and radio observations made with the VLA which showed
that any such rotation is less than 3 degrees out to redshifts in
excess of two.

This lower bound can be used for estimation of the
photon-to-Goldstone boson coupling constant $g_{a\gamma}$ using
Eq.(7) for intergalactic space. \citet{b19} and \citet{b21}
mentioned very shortly on chiral effects, including axions, that
could produced the claim effect of polarization plane rotation,
but they did not make real estimations.

We use the Eq.(7) for cosmological distances, but at first it is
necessary to explain the validity of this Eq., because by Eq.(7)
the rotation angle depends on the square of a distance $\sim L^2$.
The Eq.(7) was given only for a homogeneous magnetic field.

But the real extragalactic magnetic field changes with a distance.
For example, the magnetic field strength of the Local Super
Cluster (LSC) that is extending for $\sim (30\div 40) Mpc$,
differs of the magnetic field strength present over cosmological
distances \citep{b22}. Of course, the magnetic field structure
either of the LSC, or of cosmological distance are very poorly
known. Direct measurements through Faraday rotation give upper
limits to fields averaged along the line of sight to particular
background radio sources. The rough estimations of observational
data show that at distance scale $\sim 10 Mpc$ the magnetic field
is constrained to have $B_0\leq 10^{-9} G$, while for cosmological
distances to the present horizon $L\sim 1 Gpc$, the magnetic field
strength $B\leq 10^{-11} G$. Thus we can suggest the simple law
for the change of the intergalactic magnetic field with a distance
$B = B_0 (L_0 / L)$.

It allows us to obtain by Eq.(7) the following estimation of the
photon-to-Goldstone boson coupling constant for $z = 2$:
\begin{equation}
g_{a\gamma}\approx \frac{3.8\times 10^{-6}}{1 GeV}
\left(\frac{\theta}{3^0}\right)^{\frac{1}{2}}\left(\frac{L_0}{10
Mpc}\right)\left(\frac{10^{-15}}{B_0}\right)h_{50}
\label{eq11}
\end{equation}

\section{Some Additional Results}

\subsection{UV break in SED of QSOs}

The ultraviolet energy distribution of quasars is characterized by
a sharp steepening of the continuum shortward of the so-called
"big blue bump". The quasar "composite" spectral energy
distribution exhibits a steepening of the continuum at $\sim
1100$\AA (\citet{b45} and refs. therein). A fit of this composite
SED using a broken power-law reveals that the power-law index
changes radically from approximately (-0.69) in the near UV to
(-1.78) in the far UV. \citet{b45} label this observed sharp
steepening the "far UV break". They suggested that this break
could be caused due absorption by crystalline carbon dust. We
suggest another explanation of this observed break as being due to
the magnetic conversion of UV photons into extremely low mass
axions into the intergalactic magnetic field. It is remarkable
that the probability of such conversion process increases really
with the increase of photon energy and becomes more stronger
really in far UV range (see Eqs.(2), (3) and (4)).

Using the VLAS experimental value of $g_{a\gamma}$ amd the
characteristic length magnitude of $\sim 1 Gpc$ one can estimate
the optical depth of photon conversion into axion as
\begin{equation}
\tau\sim g_{a\gamma} B l = 3\times 10^3\times 3.8\times
10^{-15}\times 10^{-16}\times 3\times 10^{27}\approx 3.42
\label{eq12}
\end{equation}
that becomes noticeable greater that unit. It means that this
conversion process can provide the observable sharp steepening of
UV continuum of QSOs.

\subsection{Polarization of stars in the Local Bubble}

Another the intrigue problem is due to the observations of
intrinsic polarization of stars located in the Local Bubble. The
low density region of the local interstellar medium (ISM) where
the Sun is located is called the local cavity or bubble. This
region is partially filled with hot ($\sim 10^6 K$) low number
density ($\leq 0.005 cm^{-3}$) coronal gas detectable in soft
X-rays \citep{b46,b47,b48}. The Local Bubble is the interstellar
material that resides in close ($< 100 pc$) proximity to the Sun.

\citet{b49,b50} has prepared and analyzed the catalogue of optical
polarization measurements for 1000 stars closer than 50pc from the
Sun. He founded the discrepancy for a number of stars between the
measured polarization magnitude and stellar spectral
classification. It is difficult to explain the observable
polarization degree by interstellar polarization origin because of
the well known depletion of dust in the Sun's vicinity.

We suggest the process of photon conversion into extremely low
mass axions in the Local Bubble magnetic field as the probable
mechanisms of production of observable stellar polarization. Let
us estimate the characteristic oscillation lengths.

The value of plasma oscillation length is
\begin{equation}
L_P = \frac{2\pi\omega}{\omega_p^2}\cong 40\left(\frac{\omega}{3
eV}\right)\left(\frac{0.005}{N_e}\right)pc \label{eq13}
\end{equation}

The real magnetic conversion oscillation length is
\begin{equation}
L_B = \frac{2\pi}{g_{a\gamma}B} =
0.1\left(\frac{10^{-6}}{B}\right)pc
\label{eq14}
\end{equation}

Conversion theory allows to get constraints on the axion mass. The
oscillation length due mass of an axion is (see, for example,
\citet{b15}):
\begin{equation}
L_A = 20pc
\left(\frac{10^{-12}}{m_a}\right)^2\left(\frac{\hbar\omega}{3 eV
}\right)
\label{eq15}
\end{equation}

For the effective conversion photons into axions in the
interstellar medium of the Local Bubble we need
\begin{equation}
L_B\ll L_P,\,\,L_m
\label{eq16}
\end{equation}

It means that the mass of axion should be equal to $m_a <
2.6\times 10^{-11} eV$.

\subsection{The circular polarization of radiation of AGNs and QSOs}

\citet{b3} have shown in their classical work that light may
acquire a circular polarization through the effect magnetic
conversion of light into axions in the case of very small axion
masses. The magnitude of polarization acquired along the path
length $L$ in a result of magnetic conversion is determined by
Eq.(8).

At last time it is received the evidence according to which the
extragalactic objects (AGN and QSO) may show a substantial degree
of circular polarization not only in radio waves but even at high
(optical) frequencies (see \citet{b40}). Recent high resolution
observations of the quasar 3C279 with the FORS polarimeter at the
ESO-VLT indicated a variable optical circular polarization
occasionally exceeding 1\%. One can expect that appearance of
circular polarization is connected with the relativistic jets from
AGN and QSO. Recent observations made at Russian BTA-6m telescope
with the SCORPIO detector give evidence of circular polarization
of the blazar candidate S5 0716+714 at the level $\sim 0.5\%$
(Afanasiev and Amirkhanyan, private communication). A key open
question arises in the study of circular polarization, namely, on
the origin of its physical mechanism. Circular polarization in AGN
may be produced either as an intrinsic component of synchrotron
radiation or through Faraday conversion of linear polarization
into circular one. The last mechanism requires a large number of
low energy relativistic particles in the radiation region
(presumably jet or corona) to do the effective conversion and
hence a relatively low cutoff in particle energy spectrum,
approximately at the Lorentz factor $\gamma \leq 20$ \citep{b39}.
Theoretically it appears easier to generate large amounts of
circular polarization through the Faraday conversion process but
direct observational evidence is absent now (see \citet{b41}). The
key moment is the frequency dependence of observed circular
polarization that is yielded with diversity of its frequency
dependence for various objects.

Using Eq.(8) and PVLAS data for the coupling constant
$g_{a\gamma}$ one can estimate constraints on an axion mass from
polarimetric observation of AGN and QSO made by \citet{b12}.
Expected circular polarization value from $z\leq 1$ QSOs at the
level $P_V\sim 0.1\%$ gives the constraints at an axion mass as
$m_a\approx 3\times 10^{-16} eV$. That mass values lies out the
limits of standard PQ axion theory. It is curious that this axion
mass value corresponds to the magnitude obtained by \citet{b52}
from supernovae dimming.

\section{Conclusions}

Photon-pseudoscalar boson mixing birefringent effects for
cosmological distances have been considered and estimated. Effect
of cosmological alignment and cosmological rotation of
polarization for distant QSOs discovered by \citet{b12} can be
explained in terms of birefringent phenomenon due to photon - low
mass axion (Goldstone boson) mixing in a cosmological magnetic
field.

Using axion interpretation of the PVLAS data we estimate the
intergalactic magnetic field magnitude as $\sim 10^{-16} G$.

The net circular polarization of distant AGN radiation crossing an
interstellar medium can be explained as a result of magnetic
conversion of photons into extremely low mass axions. In a result
one can get the strong constraints on the nonstandard axion mass
as $m_a\leq 3\times 10^{-16} eV$.

The observations of intrinsic polarization of stars located in the
Local Bubble showed the discrepancy between the measured
polarization degree and stellar spectral classification. This fact
can't be explained by interstellar polarization. If one should
suggest the magnetic conversion process as an origin of this
discrepancy it is possible to estimate also the upper limit on an
axion mass value that appears at the level $m_a \leq 2\times
10^{-11} eV$.

\section*{Acknowledgments}

This research was made with financial support by Program of
Prezidium of Russian Academy of Sciences "Origin and Evolution of
Stars and Galaxies" and Program of the Department of Physical
Sciences of RAS "Extended Objects..." and by the LOT of Russian
Ministry of Education and Science.

\end{document}